\documentclass[twocolumn,prl]{revtex4} 
\usepackage{amssymb} 
\usepackage{graphicx}
\usepackage{subfigure}
\usepackage{dcolumn}
\usepackage{bm}
\begin{document}

\title{Structural and Electronic Properties of Oxidized Graphene}
\author{Jia-An Yan, Lede Xian, and M. Y. Chou}
\affiliation{School of Physics, Georgia Institute of Technology,
Atlanta, Georgia 30332-0430, U.S.A.}
\date{\today}

\begin{abstract}


We have systematically investigated the
effect of oxidation on the structural and electronic properties of
graphene based on first-principles calculations. Energetically
favorable atomic configurations and building blocks are identified,
which contain epoxide and hydroxyl groups in close proximity with
each other. Different arrangements of these units yield an LDA band
gap over a range of a few eV. These results suggest the possibility
of creating and tuning the band gap in graphene by varying the
oxidation level and the relative amount of epoxide and hydroxyl
functional groups on the surface.

\end{abstract}


%

\maketitle

The prospects of graphene-based nanoelectronics
\cite{DeHeer2007,Geim2007} have stimulated extensive research
activities in recent years. Pristine graphene, with two linear bands
crossing at the Dirac point, is a zero-gap material. Therefore, much
effort has been devoted to creating an energy gap in graphene-based
materials for device applications
\cite{Ohta2006,Han2007,Giovannetti2007,Li2008a,Wu2008,Jung2008,Elias2009}. In
particular, an energy gap can be achieved through either
nanopatterning \cite{Han2007,Li2008a} or chemical functionalization
\cite{Wu2008,Jung2008,Elias2009}. The latter has a greater advantage because
of the ease to scale up in the production. A recent experiment has
successfully generated a metal-semiconductor-metal junction using
epitaxial graphene and a single functionalized graphene sheet (FGS)
\cite{Wu2008}. Analogous to the combination of Si and SiO$_2$ in the
current generation of microelectronics, the FGS has the potential of
being seamlessly integrated with graphene in the fabrication of
future nanoelectronics devices. One way to produce FGSs is by
exfoliation from graphite oxide (GO), which can have different
compositions with various oxidation levels depending on the
synthesis processes and conditions. Currently, GO is of particular
interest to scientists since chemical reduction of GO has been
demonstrated as a promising solution-based route for mass production
of graphene
\cite{Schniepp2006,Stankovich2007,Gomez2007,Gilje2007,Li2008b,Eda2008}.

The electronic properties of the oxidized layers depend on the
detailed chemical structure, which remains unresolved for GO for
more than a century with various structural models proposed in the
literature
\cite{Nakjima1988,Mermoux1991,Nakajima1994,Lerf1998,He1998}.
Increased conductivity was observed during the reduction of an oxidized
graphene sheet prepared from GO \cite{Jung2008}, but the atomic origin
of this behavior is still unknown.
A recent high-resolution solid-state $^{13}$C-NMR measurement
\cite{Cai2008} has confirmed the existence of C-OH (hydroxyl), C-O-C
(epoxide), and $sp^2$ C units on these layers. The data further
indicate that a large fraction of $sp^2$ C atoms are bonded directly
to C atoms in the hydroxyl and epoxide groups, and that a large
fraction of C atoms in the hydroxyl and epoxide units are bonded to
each other. In order to fully establish the atomic configurations
and related electronic properties in this important material, we
report results from first-principles calculations that provide a
clear picture of the energetically favorable building blocks and
stable phases. Interestingly, various energy gap values are found
for structures with different O concentrations, suggesting that the
gap is highly tunable by varying the amount of hydroxyl and epoxide
on the graphene sheet.

Our theoretical study focuses on the following two key issues: (i)
how these functional groups arrange themselves on graphene, and (ii)
how these arrangements affect the electronic properties of the
graphene sheet. The calculations were carried out using the
local-density approximation (LDA) within density-functional theory
with a plane-wave basis set as implemented in the Vienna Ab-initio
Simulation Package (VASP) \cite{Kresse1993}. Vanderbilt ultrasoft
pseudopotentials \cite{Vanderbilt1990} are employed. All
calculations are done with a plane-wave cutoff energy of 500 eV.
Results shown in Fig.~1 are obtained using a 5$\times$5 graphene
unit cell with 5$\times$5$\times$1 $k$-point sampling. The
5$\times$5 unit cell contains 50 C atoms, providing a separation of
12.3~\AA~for the atomic combinations under investigation. The
vertical size of the supercell is 12~\AA, so the interaction between
the layers is expected to be minimal. The size of the lateral unit
cell is individually optimized for different coverages. The
optimization of atomic positions proceeds until the change in energy
is less than 1$\times$10$^{-6}$ eV per cell and the force on each
atom is less than 0.02 eV/\AA.


\begin{figure}[tbp]
\centering
   \includegraphics[width=7.0cm]{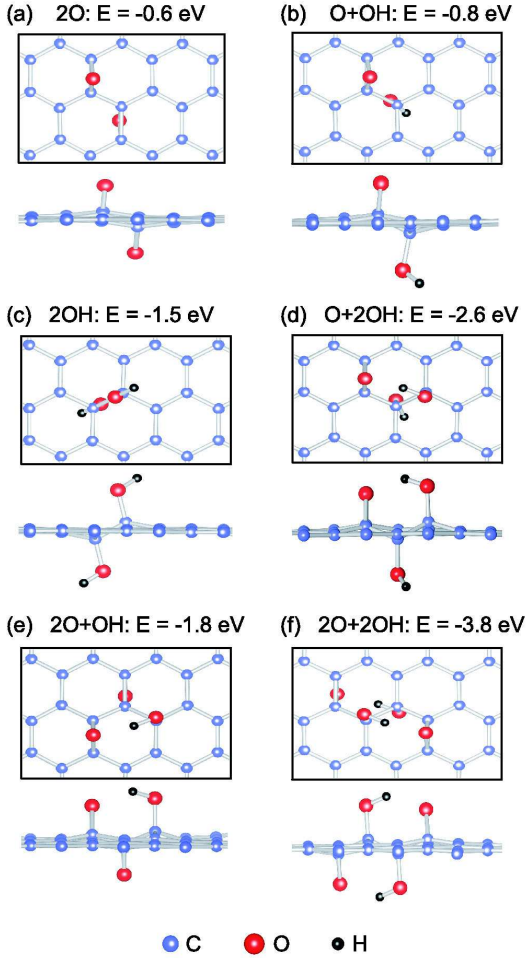}
  \caption{(Color online) Atomic configurations and energies of various favorable combinations of
epoxide and hydroxyl groups on the graphene surface. The energy
shown is calculated using a 5$\times$5 unit cell and is with respect
to the energy sum of well separated individual units on the surface.
C and O atoms are represented by large spheres and H by small
spheres. } \label{fig1}
\end{figure}


\begin{figure*}[tbp]
\centering
\includegraphics[width=11cm]{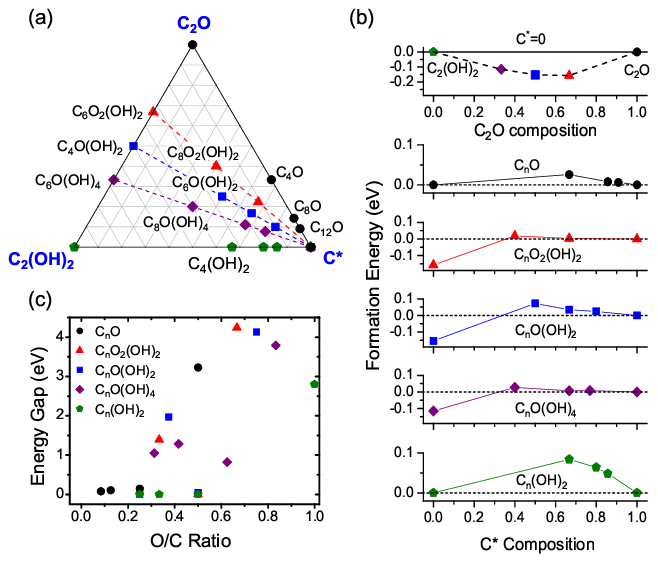}
\caption {(Color online) (a) Ternary diagram showing ordered
phases on the graphene surface with different amounts of $sp^2$
Carbon (C$^*$), epoxide (C$_2$O), and the 1,2-hydroxyl pair
(C$_2$(OH)$_2$). The phases investigated in this study are marked on
the diagram, with dashed lines indicating those with the same
relative amount of epoxide and hydroxyl pairs. (b) Formation energy
as defined in Eq.~(1) for different phases marked in the ternary
diagram in (a). (c) Energy gap as a function of the overall
oxygen-to-carbon ratio for different phases. } \label{fig2}
\end{figure*}

A single functional group of epoxide or hydroxyl on graphene can induce
significant local distortion.
With a new bond formed between C and O, the bonding characteristics of the
connecting C atoms change from planar $sp^2$ to distorted $sp^3$ hybridization.
The structure obtained in our calculation is in good agreement with that
reported in previous theoretical studies \cite{Li2006,Boukhvalov2008}.
Of particular interest is the distribution of these functional
groups on graphene. A recent atomic force microscope (AFM)
measurement showed that the oxidized graphene sheets appear to have
a thickness equal to integer multiples of $h\approx$
6.7~\AA~\cite{Pandey2008}, indicating that epoxide and hydroxyl are
most likely to be present on both sides of the graphene sheet.
Hence, all of our calculations have considered possible two-sided
configurations.

To find out whether these functional groups prefer to aggregate or
not, we have calculated the energy change associated with grouping
these epoxide and hydroxyl units in various ways, and the results
calculated using a 5$\times$5 unit cell are shown in Fig.~1. The
energies are referenced to the energy sum of separated individual
units on the surface, and the most favorable, fully-relaxed atomic
arrangements are shown in Fig.~1 for each combination. As one can
see from the figure, the energies are lowered considerably when
these epoxide and hydroxyl units are grouped together. Part of the
reason is a cancelation of vertical structural distortion when these
units can be located on both sides of the sheet. Significant energy
gains are found if the OH units form 1,2-hydroxyl pairs (Fig.~1(c))
on opposite sides of the sheet;
and the combination of O+2OH and 2O+2OH are also especially
favorable (Figs.~1(d) and (f)). In Figs.~1(d), (e), and (f), H points
toward the neighboring O on the same side, yielding a configuration
characterized by a hydrogen bond. The results in Fig.~1 conclude
that these adsorbed units prefer to aggregate on the graphene
surface. This is in full agreement with the experimental results
inferred from the NMR data \cite{Lerf1998,He1998,Cai2008}.

\begin{figure}[tbp]
\centering
\includegraphics[width=4.8cm]{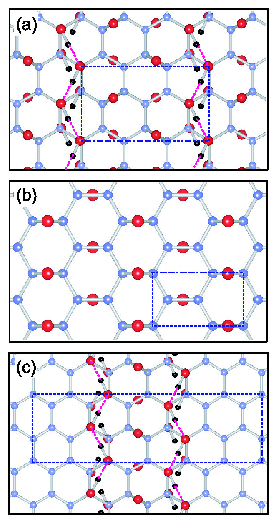}
\caption {(Color online) Atomic structures for selected phases. (a)
Fully-oxidized phase of C$_6$O$_2$(OH)$_2$.
(b) Fully-oxidized epoxide-only
phase C$_2$O with oxygen rows on both sides of the plane. (c)
C$_{24}$O$_2$(OH)$_8$ structure with hydroxyl-epoxide strips
separated by $sp^2$ carbon. C and O atoms are represented by large
spheres and H by small spheres and the dashed rectangles indicate
the respective unit cells. The hydrogen bonds in the hydroxyl chains
above the plane are indicated by dashed lines.} \label{fig3}
\end{figure}

We then explore ordered phases containing epoxide and hydroxyl
groups incorporating the energetically favorable building blocks in
Fig.~1. Based on the energy results, we only consider arrangements
with the OH group appearing in 1,2-hydroxyl pairs. Each periodic
phase can be specified by the relative amount of ``free" $sp^2$ C
atoms (denoted by C$^*$, corresponding to C atoms not bonded to O),
epoxide (C$_2$O), and the 1,2-hydroxyl pair [C$_2$(OH)$_2$ with the
connecting C atoms included]. The representative stoichiometry is
C$^*_{1-x-y}$(C$_2$O)$_x$[C$_2$(OH)$_2$]$_y$, or equivalently
C$_{1+x+y}$O$_x$(OH)$_{2y}$, with $0\leq x \leq 1$, $0\leq y \leq
1$, and $0\leq x+y \leq 1$. The ordered phases we have calculated
are marked on a ternary diagram shown in Fig.~2(a), where the dashed
lines indicate phases with the same ratio of epoxide versus the
hydroxyl pair. For each phase we investigated, the structure was
optimized by exploring various different local arrangements of the
epoxide and hydroxyl groups. The lattice parameters and atomic
coordinates are fully relaxed. The formation energy is defined in
the usual way for a ternary system:
\begin{eqnarray}
\Delta E[x,y] & = & E[{\rm C}_{1+x+y}{\rm O}_x({\rm OH})_{2y}] -
(1-x-y) E[{\rm C}^*] \nonumber \\
& &- x E[{\rm C}_2{\rm O}]- y
E[{\rm C}_2({\rm OH})_2],
\end{eqnarray}
where $E$[Z] represents the energy of a periodic phase Z. These
formation energies are shown in Fig.~2(b) for the ``binary" phases
and for phases along each dashed line in Fig.~2(a).

For the fully oxidized phases in which all C atoms are bonded to O
in either an epoxide or a 1,2-hydroxyl pair, low-energy ``binary"
phases with mixed epoxide and hydroxyl compositions are found (top
panel of Fig.~2(b)). These fully oxidized phases have a lattice
expansion of the order of 2-3 \%, and an LDA energy gap of 2.8 - 4.2
eV. With a negative formation energy, these intermediate phases are
stable against separation into pure epoxide and pure hydroxyl
phases. These stable phases include C$_6$O$_2$(OH)$_2$,
C$_4$O(OH)$_2$, and C$_6$O(OH)$_4$, with an epoxide to hydroxyl
ratio of 1:1, 1:2, and 1:4, respectively. As an example, the
structure of C$_6$O$_2$(OH)$_2$ is shown in Fig.~3(a), in which one
can easily identify the following key features: the 1,2-hydroxyl
pairs are connected to form a chain-like structure on both sides of
the sheet in such a way that the interaction associated with
hydrogen bonds can be maximized; and O atoms are drawn to the
remaining C atoms in the same hexagonal rings to form epoxides in
the close proximity. The other two fully oxidized phases,
C$_4$O(OH)$_2$, and C$_6$O(OH)$_4$ follow the same pattern for their
atomic arrangements. These features turn out to be quite
energetically favorable in constructing phases with intermediate
compositions, as will be discussed below. The epoxide-only phase,
C$_2$O, is shown in Fig.~3(b), in which the O atoms follow the
arrangements in Fig.~1(a) and stay in rows on opposite sides of the
sheet \cite{note1}. The buckling is symmetrically compensated. For
lower-concentration epoxide-only phases, the opening of the
three-membered epoxide ring by breaking the C-C bond to release
strain as proposed previously \cite{Li2006} turns out to be
energetically favorable, together with the formation of O rows.

Apart from the fully oxidized ``binaries" shown in the top panel of
Fig.~2(b), we are not able to find other ordered phases with a
negative formation energy. Therefore, the T=0 lowest-energy
configuration of the oxidized graphene sheet is likely to be a
combination of fully oxidized regions and the clean graphene phase.
This configuration has not been observed experimentally at finite
temperature. Possible reasons include the following: the entropy
term could be important at finite temperature; the oxidization
process is a highly non-equilibrium one; and the covalent bonding in
the epoxide and hydroxyl units largely reduces their mobility on the
surface. Therefore, domains of various intermediate phases may still
be found in the sample under experimental conditions. The relative
amount of epoxide and hydroxyl units on the graphene sheet depends
on the sample preparation process and can vary over a wide range.
Here we focus on phases along each dashed line in Fig.~2(a), in
which the relative amount of epoxide and hydroxyl units on the
surface is a constant. After exploring various atomic
configurations, the lowest-energy periodic structure of these
intermediate phases found in our calculation contains strips of
epoxide and hydroxyl combinations with clean graphene ribbons in
between. An example, the structure of C$_{24}$O$_2$(OH)$_8$, is
shown in Fig.~3(c), which contains separate regions of $sp^2$ and
$sp^3$ carbon. The $sp^3$ strips consist of double hydroxyl chains
and neighboring epoxides. The formation energies for the periodic
phases with various intermediate compositions constructed in this
fashion are shown in Fig.~2(b). The strips interact weakly when
separated by carbon ribbons, which explains why the formation
energies of these phases fall on a straight line in Fig.~2(b) when
the C$^*$ composition is larger than about 0.4. (Phases with
different C$^*$ compositions have different strip separations.) With
the formation energies falling on a straight line, many of these
phases are expected to coexist on the surface. It is energetically
favorable for these strips to coalesce, since the fully oxidized
phase has a lower formation energy. However, this process may not be
completed during the oxidation process.

The current results indicate that the configurations investigated
in previous density-functional calculations \cite{Boukhvalov2008}
for oxidized graphene sheets containing both the epoxy and hydroxyl groups
may not be energetically favorable for a given composition.
In our calculation, we find that the formation of hydroxyl chains
resulting from hydrogen-bond interaction between neighboring 1,2-hydroxyl
pairs greatly lowers the energy and that the epoxides are grouped next
to these hydroxyl chains. The epoxide and hydroxyl units randomly
deposited on the surfaces are expected to arrange themselves locally
following these preferred patterns, giving rise to patches of $sp^2$
carbon surrounded by fully oxidized epoxide+hydroxyl regions or vice
versa. The possible existence of pure graphene ribbons was proposed
previously \cite{Kudin2008} in order to explain the shift of the
Raman peaks in GO and FGSs. The current study provides a strong
support for this picture based on extensive first-principles
calculations.

The finite region of $sp^2$ carbon has a consequence in the
electronic structure. Without knowing the exact atomic arrangements
at various compositions, we use the results of the ordered
structures investigated above to provide an estimate for this
effect. The calculated LDA energy gaps associated with the ordered
phases considered in Fig.~2(a) as a function of the overall O/C
ratio are shown in Fig.~2(c). The data-point symbols are the same as
those in other parts of Fig.~2.
An LDA energy gap of the size ranging from zero to 4
eV can be found for the range of O concentration we considered. The
gap range includes both semiconducting and insulating phases, and
the gap size is dictated by the width of the graphene ribbons in the
ordered phases we studied. (It is well known that an LDA calculation underestimates the energy gap, and the real gap value is expected to be larger.) A few vanishing band gap values are
associated with armchair ribbons with 3$n$+2 rows of atoms ($n$ is
an integer) or zigzag ribbons with an even number of atomic rows.
These may be considered as special cases. The overall band-gap
results suggest a promising and practical way to tune the energy gap
in FGSs by varying the degree of oxidization, which is feasible
experimentally, and possibly the location of the oxidized regions.

In summary, with first-principles calculations we have studied the
structure and energetics of oxidation functional groups (epoxide and
hydroxyl) on single-layer graphene, and the induced changes in the
electronic properties. We find that it is energetically favorable
for the hydroxyl and epoxide groups to aggregate together and to
form specific types of strips with $sp^2$ carbon regions in between.
An LDA band gap ranging from a few tenths of an eV to 4 eV can be
obtained by changing the oxidation level and the location of the
oxidized region, suggesting a great potential to tune the energy gap
in graphene through controlled oxidation processes.

We acknowledge stimulating discussions with W. de Heer, C. Berger,
X. Wu, and M. Sprinkle. J.A.Y. thanks D. Pandey for sending the
preprint of their paper. This work is supported by the Department of
Energy (Grant No. DE-FG02-97ER45632). L. X.  acknowledges support from the Georgia Tech MRSEC funded by the
National Science Foundation (Grants No. DMR-08-20382). This research used computational resources at the
National Energy Research Scientific Computing Center, which is
supported by the Office of Science of the U.S. Department of Energy
under Contract No. DE-AC02-05CH11231, and the National Science
Foundation TeraGrid resources provided by the Texas Advanced
Computing Center (TACC).


\end{document}